%% file: prper_phys_SEM.tex
\documentclass[aps,prl,twocolumn,groupedaddress]{revtex4-1}

\pdfoutput=1

\usepackage{graphicx} 
\usepackage{enumitem} 
\usepackage{subcaption} 
\usepackage{multirow} 

\begin{document}

\title{In a physics curriculum only introductory physics course grades show gender differences but do not predict future course performance for physics majors}

\author{Kyle~M.~Whitcomb}
\affiliation{Department of Physics and Astronomy, University of Pittsburgh, Pittsburgh, PA, 15260}
\author{Chandralekha~Singh}
\affiliation{Department of Physics and Astronomy, University of Pittsburgh, Pittsburgh, PA, 15260}

\date{\today}

\begin{abstract}
	Analysis of institutional data for physics majors showing predictive relationships between required mathematics and physics courses in various years is important for contemplating how the courses build on each other and whether there is need to make changes to the curriculum for the majors to strengthen these relationships.
	We use 15 years of institutional data at a large research university to investigate how introductory physics and mathematics courses predict male and female physics majors' performance on required advanced physics and mathematics courses.
	We used Structure Equation Modeling (SEM) to investigate these predictive relationships and find that among introductory and advanced physics and mathematics courses, there are gender differences in performance in favor of male students only in the introductory physics courses after controlling for high school GPA.
	We found that a measurement invariance fully holds in a multi-group SEM by gender, so it was possible to carry out analysis with gender mediated by introductory physics and high school GPA.
	Moreover, we find that these introductory physics courses that have gender differences do not predict performance in advanced physics courses.
	Also, introductory mathematics courses predict performance in advanced mathematics courses which in turn predict performance in advanced physics courses.
	Furthermore, apart from the introductory physics courses that do not predict performance in future physics courses, there is a strong predictive relationship between the sophomore, junior and senior level physics courses.
\end{abstract}

\maketitle

\section{Introduction and Theoretical Framework}

The importance of evidence-based approaches to improving student learning and ensuring that all students have the opportunity to excel regardless of their background is becoming increasingly recognized by physics departments across the US~\cite{henderson2008, dancy2010, henderson2012}.
Holistic consideration of how these physics departments are currently succeeding in supporting their undergraduate majors is crucial in order to make appropriate changes to the curricula and pedagogies for the majors based upon metrics informed by data and ensure that students are adequately supported and advised.
These considerations include how prerequisite physics and mathematics courses predict performance in subsequent physics courses throughout the curriculum for physics majors, and such investigations are vital regardless of the theory of change~\cite{henderson2008, dancy2010} a physics department adopts and implements based upon its institutional affordances and constraints.
At the same time, with advances in digital technology in the past decade, data analytics can provide valuable information that can be useful in transforming learning for all students~\cite{baker2014, papamitsiou2014}.

The theoretical framework for this research is inspired by the fact that while many investigations in physics education have focused on evidence-based classroom practices to improve student learning at all levels in the physics undergraduate curriculum~\cite{mcdermott1987, mcdermott1995, singh2008, lin2013, zhu2012, zhu2013, sayer2016}, there is significantly less focus on the connection between how student performance in different subsequent courses builds on prior courses in the physics curriculum overall.
Information obtained from data analytics on large institutional data in these areas can be an important component of understanding, e.g., the role the earlier courses play in later physics course performance as well as contemplating strategies for strengthening these ties, and improving physics major advising, mentoring and support.
Moreover, it is important that physics departments take a careful look at the extent to which their programs for the majors are equitable and inclusive and provides adequate support, advising and mentoring to all students, including women and students from diverse ethnic and racial backgrounds who have traditionally been left out in order to ensure that all students have sufficient opportunity to excel as a physics major.

In order to gain an understanding of these issues central to improving physics education for the majors, this research harnesses data analytics in the context of a large state-related university to investigate how well the performance of physics majors in physics and mathematics courses throughout a physics curriculum predicts performance in subsequent physics courses.
We note that the first-year physics and mathematics courses are very similar at most colleges in the US.
Moreover, many of the advanced courses in these subjects also have well-defined curricula that are common across many colleges and universities.
These courses for the majors have been offered over decades under the assumption that the later physics courses would build on the earlier ones coherently to help the majors build a robust knowledge structure of physics and develop their problem solving, reasoning and meta-cognitive skills.
Here we discuss an investigation that uses data analytics applied to 15 years of institutional data for physics majors to analyze not only these relationships between course performance in different years, but also whether there are any gender differences in these curricular relationships.
The investigation can be useful for other institutions who may perform similar analyses in order to contemplate strategies for improving education for physics majors in a holistic manner.
In particular, institutions could compare their findings with the baseline data from a large state-related university presented here for the synergy observed between the required courses in the curriculum for the physics major.

\section{Research Questions}

Our research questions regarding the physics curriculum for the majors at a large state-related university are as follows.

\begin{enumerate}[label={\bfseries RQ\arabic*.}, ref={\bfseries RQ\arabic*}, itemsep=1pt]
	\item \label{rq_courses} Are there gender differences in course performance among physics majors in introductory and advanced physics and mathematics courses?
	\item \label{rq_sem} Does performance in introductory physics and mathematics courses predict performance in advanced physics and mathematics courses?
	\item \label{rq_gender} Does the degree to which earlier course grades predict later course grades differ for men and women?
\end{enumerate}

\section{Methodology}

\subsection{Measures}

\begin{table}
	\input{tex/table_curriculum}
	\caption{\label{table_courses} All required lecture courses in physics and mathematics taken by physics majors are listed.
	Full course names are given along with shortened names used elsewhere in this paper and the term(s) in which the courses are typically taken by physics majors.}
\end{table}

Using the Carnegie classification system, the university at which this study was conducted is a public, high-research doctoral university, with balanced arts and sciences and professional schools, and a large, primarily residential undergraduate population that is full-time and reasonably selective with low transfer-in from other institutions~\cite{carnegie}.
De-identified data were provided by the university on all students who had enrolled in introductory physics from Fall 2005 through Spring 2019.
The data include demographic information such as gender.
We note that gender is not a binary construct.
However, the university data includes ``gender'' as a binary categorical variable.
Therefore, that is how the data regarding gender are represented in these analyses.
From the full sample from 2005-2019, a sub-sample was obtained by applying several selection criteria to select out physics majors from those from other majors who took introductory physics.
In particular, in order to be kept in the sample, students were required to meet the following criteria:
	1) declare a physics major at any point or be a non-engineering student enrolled in the honors introductory sequence,
	2) enroll in at least 30\% of the courses listed in Table~\ref{table_courses}, and
	3) enroll in Modern Physics.
Note that all of the courses we consider in this analysis in Table~\ref{table_courses} are the required lecture courses in the curriculum for the physics major.
We consider only required courses in order to maintain as consistent a population as possible.
Further, we consider only lecture courses since the contemporary laboratory courses have very high and narrow grade distributions (with over 90\% of students receiving an A) that are less suited for investigations of gender differences.
After applying the selection criteria, the sample contains 451 students, which are 19.5\% female and have the following race/ethnicities: 80.5\% White, 10.9\% Asian, 2.4\% Latinx, 2.2\% African American, and 3.8\% Other or Unspecified.

The data also include high school GPA on a weighted 0-5 scale that includes adjustments to the standard 0-4 scale for Advanced Placement and International Baccalaureate courses.
Further, students' declared majors are reported separately for each term in which they are enrolled.
Finally, the data include the grade points and letter grades earned by students in each course taken at the university.
Grade points are on a 0-4 scale with $\text{A}=4$, $\text{B}=3$, $\text{C}=2$, $\text{D}=1$, $\text{F}=0$, where the suffixes `$+$' and `$-$' respectively add or subtract $0.25$ grade points (e.g. $\text{B}-=2.75$), with the exception of $\text{A}+$ which is reported as the maximum 4 grade points.

\subsection{Analysis}

In order to evaluate the grades that the physics majors earn in physics and mathematics courses, we grouped students by the gender variable and computed standard descriptive statistics (mean, standard deviation, sample size) separately for each group.
Gender differences in course grades were initially evaluated using Cohen's $d$ to measure the effect size~\cite{neter1996, montgomery2012}, as is common in education research~\cite{nissen2018}.

The extent to which performance (i.e., grades earned) in earlier physics and mathematics courses predicts performance in later physics and mathematics courses was evaluated using Structural Equation Modeling (SEM)\cite{kline2011}.
SEM is the union of two statistical modeling techniques, namely Confirmatory Factor Analysis (CFA) and Path Analysis.
The CFA portion tests a model in which observed variables (or ``indicators'') are grouped into latent variables (or ``factors''), constructed variables that represent the variance shared among all indicators that load on that particular factors.
The degree to which each indicator is explained by the factor is measured by the standardized factor loadings, $\lambda$ (with $0 \leq \lambda \leq 1$), where $\lambda^{2}$ gives the percentage of variance in the indicator explained by the factor.
The Path Analysis portion then tests for the statistical significance and strength of regression paths between these factors, simultaneously estimating all regression coefficients, $\beta$, throughout the model.
This is an improvement over a multiple linear regression model in which only a single response (target or outcome) variable can be predicted at a time, which problematically disallows hierarchical structures~\cite{dusen2019}.
By estimating all regression paths simultaneously, all estimates are able to be standardized simultaneously, allowing for direct comparison between standardized $\beta$ coefficients throughout the model.

In this paper, we report the model fit for SEM using the Comparitive Fit Index (CFI), Tucker-Lewis Index (TLI), and Root Mean Square Error of Approximation (RMSEA).
Commonly cited standards for goodness of fit using these indices are as follows:
For CFI and TLI, Hu and Bentler~\cite{hu1999} found that many authors~\cite{hu1999, carlson1993, rigdon1996} suggest values above 0.90 and 0.95 indicate a good fit and a great fit, respectively.
For RMSEA, several authors \cite{hu1999, browne1993} suggest that values below 0.10, 0.08, and 0.05 indicate a mediocre, good, and great fit, respectively.

Finally, these model estimations can be performed separately for different groups of students (e.g., men and women) using multi-group SEM.
These differences are measured in a series of tests corresponding to different levels of ``measurement invariance'' in the model~\cite{kline2011}, with each step fixing different elements of the model to equality across the groups and comparing to the previous step via a Likelihood Ratio Test (LRT).
A non-significant $p$-value at each step indicates that the estimates are not statistically significantly different across groups.
``Weak'' measurement invariance is demonstrated by fixing to equality the factor loadings, ``strong'' invariance is demonstrated by further fixing to equality the indicator intercepts, and finally ``strict'' invariance is demonstrated by further fixing to equality the residual error variance of the indicators.
If measurement invariance holds, then all remaining differences between the groups occur at the factor level, either as differences in factor intercepts or $\beta$ coefficients.
Further, if no differences are found in $\beta$ coefficients, then any remaining group differences in factor intercepts may be modeled by including a categorical grouping variable which directly predicts the factors.

Using SEM, we model student progression through the physics curriculum by grouping courses together into factors by their subject (physics or mathematics) and the order in which the courses are typically taken by physics majors.
We use multi-group SEM to test for gender moderation, i.e., to test for gender differences in the model, including mean differences of courses (indicators) and course factors.
Since we found no gender differences anywhere except in factor-level intercepts, we ultimately model the gender differences not with multi-group SEM, but with a categorical ``Gender'' variable directly predicting items with different intercepts.

Due to the nature of institutional grade data, modeling students' progress through an entire curriculum involves a large amount of missing data due to various factors.
These can include students receiving credit for courses taken elsewhere (e.g., over the summer at a different college), not completing the curriculum, skipping courses that are normally required with special permission, and the inevitable errors that occur in large datasets.
The default approach to missing data in many modeling programs, listwise deletion, is then not desirable since it leaves very few students in the sample and can bias the results~\cite{nissen2019}.
Considering this, we employed Full Information Maximum Likelihood (FIML) in order to impute missing data within the SEM model~\cite{kline2011}.

In addition to the aforementioned benefits of using SEM such as simultaneous estimation of all model elements and the ability to use FIML for missing data estimation, the basic structure of SEM also provides benefits to the modeling process.
In particular, by first using CFA to group indicators into factors and then performing path analysis on those factors, the effect of measurement error is minimized since the error variance will be left at the indicator level and does not contribute to the estimation of regression coefficients at the factor level~\cite{kline2011}.

All analyses were conducted using R~\cite{rcran}, making use of the package \texttt{lavaan}~\cite{lavaan} for the SEM analysis and the package \texttt{tidyverse}~\cite{tidyverse} for data manipulation and descriptive statistics.

\section{Results}


\begin{table}[t]
\input{tex/table_introphys}
\caption{\label{table_introphys}
	Descriptive statistics are reported for prospective physics majors in introductory physics courses.
	To be included in this table, students need only have declared a physics major, not necessarily enrolled in advanced courses, in order to briefly examine all students who declared a physics major during their first year.
	The reported statistics include the sample size ($N$), mean grade points earned ($\mu$), and standard deviation of grade points ($\sigma$) in each course for men and women separately, along with Cohen's $d$ measuring the effect size of the gender difference.
	$d < 0$ indicates the mean for men is higher, $d > 0$ indicates the mean for women is higher.
}
\end{table}

\begin{table}[h!]
\input{tex/table_physmajors}
\caption{\label{table_physmajors}
	Descriptive statistics are reported for physics and mathematics courses taken by physics majors who have at least taken physics courses up to and including Modern Physics.
	Reported are the sample size ($N$), mean grade points earned ($\mu$), and standard deviation of grade points ($\sigma$) in each course for men and women separately, along with Cohen's $d$ measuring the effect size of the gender difference.
	$d < 0$ indicates the mean for men is higher, $d > 0$ indicates the mean for women is higher.
	Three multivariate analyses of variance (MANOVA) are reported, with courses grouped to reduce listwise deletion into introductory physics, advanced physics, and mathematics.}
\end{table}

In order to investigate for gender differences in course grades and answer \ref{rq_courses}, we grouped students by the gender variable and first calculated the standardized mean difference, Cohen's $d$, to measure the effect size of the gender differences~\cite{neter1996, montgomery2012}.
Table~\ref{table_introphys} shows these results for the required physics and mathematics courses for prospective physics majors in their first year courses, regardless of whether they continued on in the curriculum, while Table~\ref{table_physmajors} shows these results for only those who at least continued through Modern Physics.
Though all later analyses are performed on the student population shown in Table~\ref{table_physmajors}, namely those physics majors who persist at least through the second year, the contrast between those students and the first-year prospective physics in Table~\ref{table_introphys} shows that on average higher than average performing women in the Honors Physics courses are switching out of a physics major.
Note that some courses have lower $N$ than Modern Physics for a variety of reasons such as skipping the course with Advanced Placement credit (Physics 1, Calculus 1, Calculus 2), the course not always being required for the major (Comp. Methods, Thermo, QM 1), or students either having requirements waived or obtaining credit at other universities (potentially all courses).

We find that, on average, men performed slightly better than women in all introductory physics courses, with Cohen's $d$ ranging from $-0.12$ to $-0.24$ among all prospective physics majors (Table~\ref{table_introphys}), indicating a small effect size, and ranging from $-0.16$ to $-0.49$ for those who continue to Modern Physics (Table~\ref{table_physmajors}), indicating a small to medium effect size.
Gender differences in mathematics and advanced physics courses (Table~\ref{table_physmajors}) are mixed, with no clear pattern of performance differences.

The statistical significance of these gender differences is first tested using a multivariate analysis of variance (MANOVA) on three clusters of courses in Table~\ref{table_physmajors}, namely introductory physics, advanced physics, and mathematics.
Courses were clustered in order to keep the number of students in the MANOVA from dropping too low, since MANOVA employs listwise deletion.
These results support the patterns noted before: that introductory physics ($F(2,371)=3.13$, $p=0.045$) displays a consistent pattern of men earning higher grades than women, albeit only marginally significant at the $p<0.05$ level with the listwise deletion employed by MANOVA.
Further, there is no consistent pattern in either advanced physics ($F(6,119)=1.52$, $p=0.179$) or advanced mathematics ($F(5,108)=1.07$, $p=0.379$), evidenced by $p > 0.05$ for each of these tests.
A more sophisticated test of these gender differences will occur in the investigation of \ref{rq_gender}, where we can use multi-group SEM to test for gender differences among all elements of the model, including differences in the means earned by men and women in each course.
In addition, multi-group SEM allows us to perform these tests while using FIML to estimate missing data, a significant improvement over listwise deletion.


\begin{figure*}[htp]
	\includegraphics{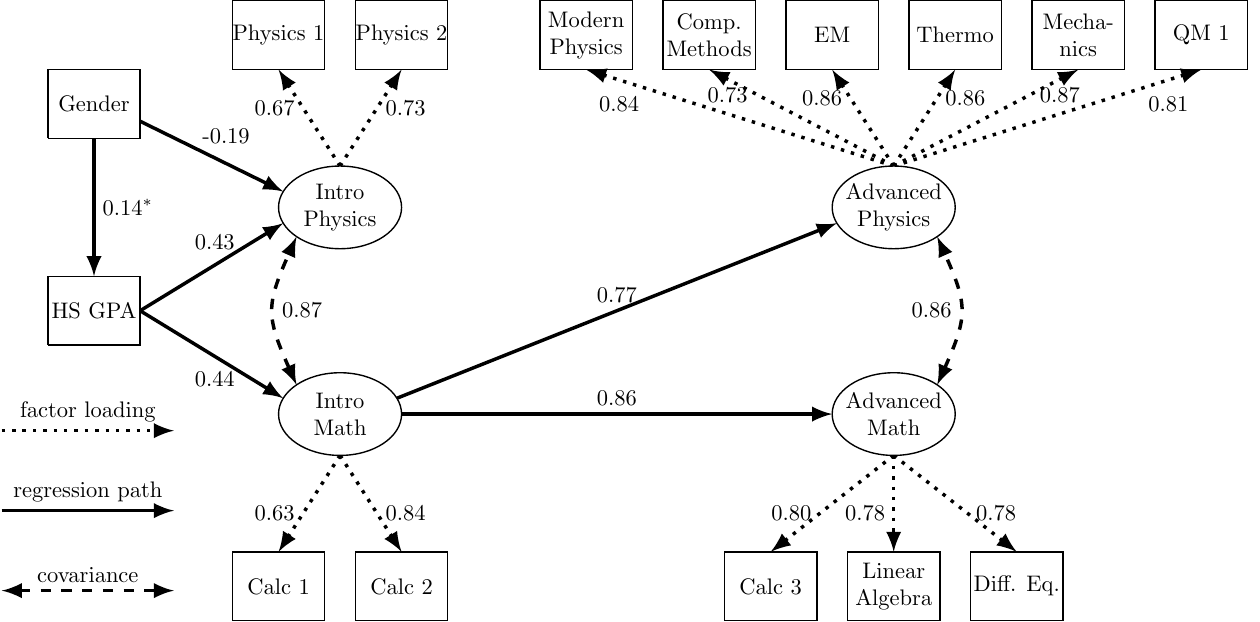}
	\caption{
		\label{figure_semA}
		A diagram of the SEM model designed to test for the relationship between physics and mathematics courses in the physics curriculum, as well as gender differences therein.
		Reported next to each line are the standardized values for factor loadings, regression coefficients, and covariances.
		The gender variable was coded as 1 for ``F'' and 0 for ``M'', so paths from gender with $\beta > 0$ and $\beta < 0$ indicate a higher mean for women and men, respectively, in the predicted variable.
		All drawn paths are significant to the $p<0.001$ level except those denoted with a superscript $*$, which are significant to the $p<0.01$ level.
		All missing paths are not statistically significant, with $p > 0.05$.
	}
\end{figure*}

\begin{figure*}[htp]
	\includegraphics{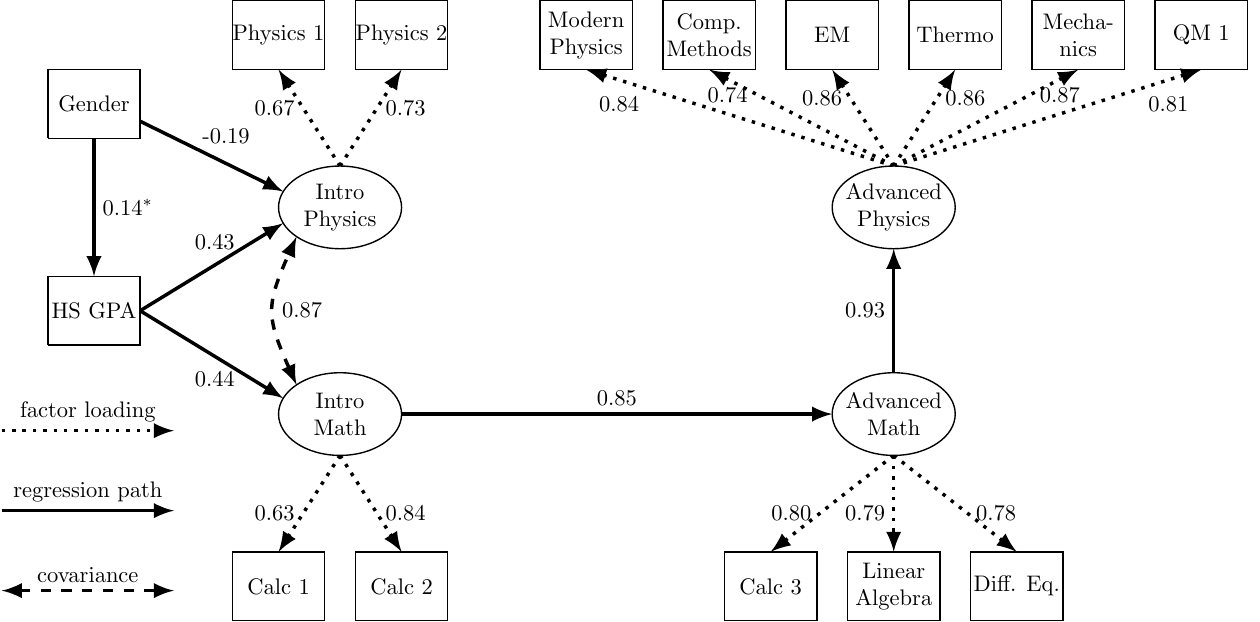}
	\caption{
		\label{figure_semB}
		An alternate model to the one shown in Fig.~\ref{figure_semA}, with the Advanced Math factor allowed to predict Advanced Physics.
		Reported next to each line are the standardized values for factor loadings, regression coefficients, and covariances.
		The gender variable was coded as 1 for ``F'' and 0 for ``M'', so paths from gender with $\beta > 0$ and $\beta < 0$ indicate a higher mean for women and men, respectively, in the predicted variable.
		All drawn paths are significant to the $p<0.001$ level except the one denoted with a superscript $*$, which is significant to the $p<0.01$ level.
		All missing paths are not statistically significant, with $p > 0.05$.
	}
\end{figure*}

\begin{figure*}[htp]
	\includegraphics{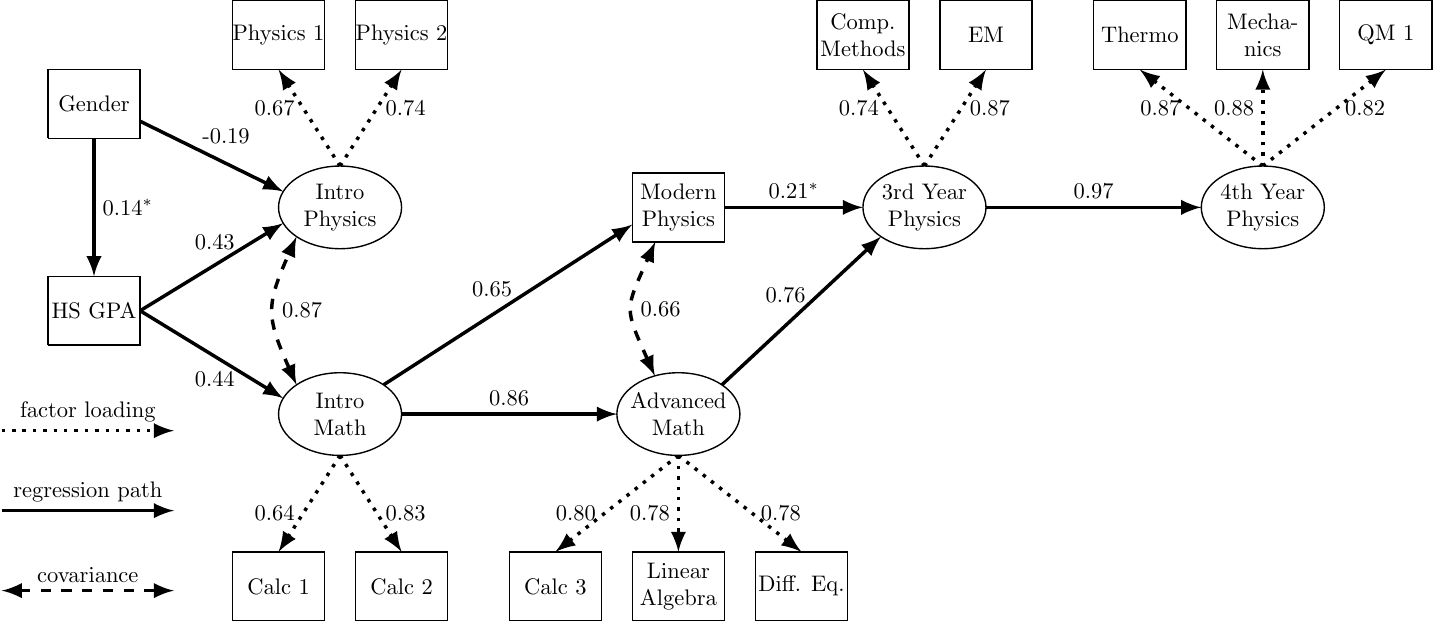}
	\caption{
		\label{figure_semC}
		An alternate model to the one shown in Fig.~\ref{figure_semA}, with the Advanced Physics factor split by the year in which the courses are typically taken.
		Reported next to each line are the standardized values for factor loadings, regression coefficients, and covariances.
		The gender variable was coded as 1 for ``F'' and 0 for ``M'', so paths from gender with $\beta > 0$ and $\beta < 0$ indicate a higher mean for women and men, respectively, in the predicted variable.
		All drawn paths are significant to the $p<0.001$ level except the one denoted with a superscript $*$, which is significant to the $p<0.01$ level.
		All missing paths are not statistically significant, with $p > 0.05$.
	}
\end{figure*}

Turning then to \ref{rq_sem}, we use SEM to test for the degree to which performance in earlier courses predicts that of later courses in the curriculum.
The full $451$ student sample was used in all SEM models, with FIML employed to impute missing data.
We grouped courses into four broad categories: introductory physics (with the regular and honors sequences combined), advanced physics (all physics beyond the introductory sequence), and introductory mathematics (Calculus 1 and Calculus 2), and allowed regression paths forward in time from introductory to advanced courses.

The final model is shown in Fig.~\ref{figure_semA} ($\text{CFI} = 0.947$, $\text{TLI} = 0.933$, $\text{RMSEA} = 0.053$), in which non-significant regression paths have been trimmed from the model.
One notable feature of Fig.~\ref{figure_semA} is that introductory mathematics strongly predicts advanced mathematics, as expected, which covaries strongly with advanced physics.
However, introductory physics does not predict advanced physics at all while introductory mathematics does, indicating that the primary predictor of success in advanced physics courses is success in mathematics courses.

Figure~\ref{figure_semA} is not the only possible model for the relationships among courses.
In particular, the majority of students take all of the advanced mathematics courses either before or concurrently with all advanced physics courses beyond Modern Physics.
A model in which advanced mathematics predicts rather than covaries with advanced physics is shown in Fig.~\ref{figure_semB} ($\text{CFI} = 0.946$, $\text{TLI} = 0.934$, $\text{RMSEA} = 0.053$).
Yet another model is shown in Fig.~\ref{figure_semC} ($\text{CFI} = 0.950$, $\text{TLI} = 0.936$, $\text{RMSEA} = 0.052$), in which the advanced physics factor has been split according to the typical time-order in which students take the courses.
No models tested show introductory physics predicting advanced physics when controlling for introductory and/or advanced mathematics, including those not shown here such as a model in which introductory mathematics is allowed to predict introductory physics, rather than covary with it.


To test for gender differences and answer \ref{rq_gender}, we first used multi-group SEM to estimate the model separately for men and women, and then used a series of likelihood ratio tests to test for differences in the model~\cite{kline2011}, first testing factor loadings, then indicator intercepts, then residual variances, then finally regression paths.
In each step, the model fit was moderate to good, with $\text{CFI} > 0.90$, $\text{TLI} > 0.90$, and $\text{RMSEA} < 0.08$.
Each step produced statistically non-significant changes from the previous according to LRTs, indicating that the estimates could be fixed to equality across the two groups ($p > 0.10$ for each step).
The only statistically significant gender differences occurred in the intercepts of high school GPA, which is not an indicator for any factor, and the introductory physics factor.
Since there were no statistically significant gender differences in regression coefficients, we converted the model from a multi-group SEM to a model that includes gender as a binary categorical variable (1 for ``F'' and 0 for ``M'') predicting high school GPA and introductory physics.

In all three models (Figs.~\ref{figure_semA}, \ref{figure_semB}, and \ref{figure_semC}) the gender differences take on the same form: on average, women have slightly higher high school GPA ($\beta = 0.14$, $p = 0.002$), while men are predicted to have slightly higher grades in introductory physics ($\beta = -0.19$, $p < 0.001$) when controlling for the high school GPA difference, and no other gender differences are predicted anywhere else in the model.
To expand further, the statistically significant path from gender to introductory physics means that men are predicted to have higher grades in introductory physics than women with the same high school GPA.
For the courses other than introductory physics, this means that the inconsistent gender differences observed in mathematics and advanced physics courses in Table~\ref{table_physmajors} are statistically non-significant when controlling for high school GPA, which either directly or indirectly predicts every other course in the model.

\section{Discussion}

In answering each of the research questions, the introductory physics sequence stood out as behaving differently from the other courses, and the overall picture paints the introductory sequence as the only gender-imbalanced part of the entire physics curriculum (pertaining to differential performance of men and women).
In particular, answering~\ref{rq_courses}, Tables~\ref{table_introphys} and~\ref{table_physmajors} together with the gender differences observed in the SEM models in Figs.~\ref{figure_semA}, \ref{figure_semB}, and \ref{figure_semC} show that introductory physics courses are the only ones in the curriculum with statistically significant gender differences, with men earning higher grades on average than women.
The SEM models provide further context, showing that all other gender differences are non-significant when controlling for high school GPA, which is higher on average for female physics majors than their male counterparts.
Thus, even though men only earn higher grades in introductory physics with a small effect size, that small effect size is slightly larger in magnitude and opposite in sign to the effect size of women's higher average high school GPA.
One hypothesis for why there is gender difference in performance in introductory courses is that those courses are taken in the first year in large-enrollment classes.
Due to societal stereotypes and biases associated with physics, women may have a lower sense of belonging and self-efficacy in those types of impersonal, non-equitable, and non-inclusive learning environments which can impact learning.

In answering \ref{rq_sem}, the SEM model in Fig.~\ref{figure_semA} shows that performance in introductory physics does not predict future grades earned in advanced physics courses when controlling for performance in introductory mathematics, and this is true for both men and women.
We also note that whether we allow advanced mathematics to covary with advanced physics (Fig.~\ref{figure_semA}) or predict advanced physics directly (Fig.~\ref{figure_semB}), we find no statistically significant regression path from introductory to advanced physics.
However, allowing advanced mathematics to predict (via a regression path) rather than covary with advanced physics leads to advanced physics being predicted solely by advanced mathematics (and not by introductory mathematics).
That is, in Fig.~\ref{figure_semB}, introductory mathematics strongly predicts advanced mathematics, which in turn strongly predicts advanced physics.
One reason for why introductory mathematics only predicts advanced physics via advanced mathematics (Fig.~\ref{figure_semB}) is that the content of Calculus 1 and 2 courses (e.g., evaluating limits and simple differentiation and integration) is less directly relevant to success in advanced physics courses.
While one is expected to know simple differentiation and integration in advanced physics courses, most of the variance in advanced physics performance is due to student proficiency in vector calculus, linear algebra and differential equations (in fact, in these physics courses, students generally get full credit for leaving the final answer as an integral if the limits and integrand are correct).

Further, Fig.~\ref{figure_semC} explores the relationship among future physics courses and finds statistically significant regression paths from Modern Physics (the sole required 2nd year physics course) to 3rd year physics to 4th year physics, even when controlling for advanced mathematics.
Yet still, Fig.~\ref{figure_semC} shows no connection from introductory physics to any other courses.
This makes the lack of a connection from introductory physics to future physics courses unique in the physics sequence.

One hypothesis for why only advanced mathematics courses predict performance in advanced physics courses while introductory physics courses do not is that advanced physics courses essentially test student facility with mathematical procedures as opposed to their conceptual understanding which is typically the focus in introductory physics courses.
In particular, students can typically do very well in advanced physics courses if they have just enough knowledge of advanced physics in order to recognize which mathematical procedure to use (e.g., solving a boundary value problem) even if their conceptual foundation in physics is weak (which is the focus of introductory physics).
In fact, our earlier investigation pertaining to conductors and insulators suggests that advanced physics students on average do not perform better on conceptual questions at the level of introductory physics than introductory physics students~\cite{bilak2007}.
Moreover, in another investigation, many students in advanced graduate courses did not perform significantly better than introductory students and admitted that they had no time to think about concepts and were essentially solving mathematics problems without learning physics from their advanced courses~\cite{maries2013}.

Finally turning towards \ref{rq_gender}, we find that in all three SEM models tested (Figs.~\ref{figure_semA}, ~\ref{figure_semB}, and ~\ref{figure_semC}), there were no significant gender differences in any predictive paths in the model.
The gender differences only occurred in two places: the intercepts of high school GPA (higher on average for women) and the introductory physics factor (higher on average for men).
Introductory physics did not predict forward at all in the SEM model, but high school GPA predicts every course factor in the model either directly or indirectly.
This means that any gender differences elsewhere in the model (i.e., not in introductory physics) are consistent with those observed in high school GPA.

We note that this model focuses on the relationships between the grades earned and does not account for other ways in which gender disparities in introductory physics can affect students (e.g., through self-efficacy, sense of belonging, etc.).
However, grades earned play a key role in students' crucial decisions about whether to remain in college and which major to pursue~\cite{betz1986, wang2013, lichtenberger2013, zimmerman2000}.
In particular, one mechanism by which this occurs is the feedback loop between course grades and self-efficacy~\cite{zimmerman2000, bandura1991, bandura1994, bandura1997, pintrich2003, pajares2005, britner2006}.
Other studies at this same university have found significant gender differences favoring men in the physics self-efficacy of students in large introductory physics courses~\cite{marshman2017, marshman2018}, consistent with studies at other universities~\cite{cavallo2004, sawtelle2012}.
Although our analysis only includes students who had not only declared a physics major but had also completed the modern physics course in the sophomore year, the gender differences in grades earned in introductory physics courses could have a large, gender differential impact on students' choice to pursue physics and related majors, despite the fact that performance in introductory physics does not predict performance in advanced physics courses.
These findings further suggest the need for efforts towards improving equity and inclusion in introductory physics courses, including interventions designed to boost students' self-efficacy, growth mindset and sense of belonging in physics~\cite{wilson1985, walton2007, walton2018, binning}.

In conclusion, a completely cohesive curriculum for physics majors should not only be consistent in academic content from year to year, but also in its positive and inclusive environment so that students from all demographics can excel including those groups which have traditionally been underrepresented in physics.
We urge researchers at other institutions to perform similar analyses in order to evaluate the efficacy of the assumptions underlying the curriculum for physics majors, and how well the various courses required for physics majors cohere.

\section{Acknowledgments}
This research is supported by the National Science Foundation Grant DUE-1524575 and the Sloan Foundation Grant G-2018-11183.
We thank Robert Devaty for his invaluable feedback on this manuscript.

\bibliography{refs}

\appendix
\section{Appendix: Grade Distributions by Gender}

\begin{figure*}
    \centering
	\begin{subfigure}{0.475\textwidth}
		\includegraphics[width=\linewidth]{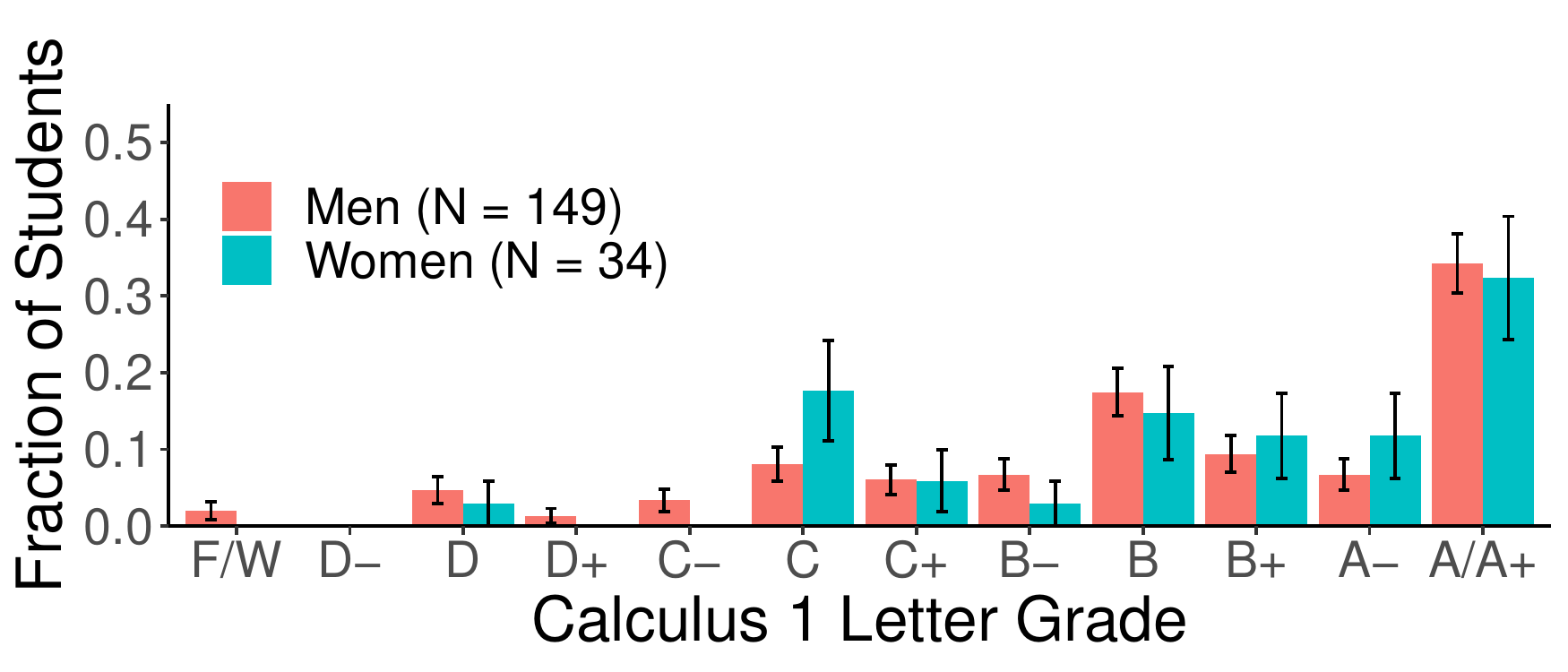}
	\end{subfigure}\hfil
	\begin{subfigure}{0.475\textwidth}
		\includegraphics[width=\linewidth]{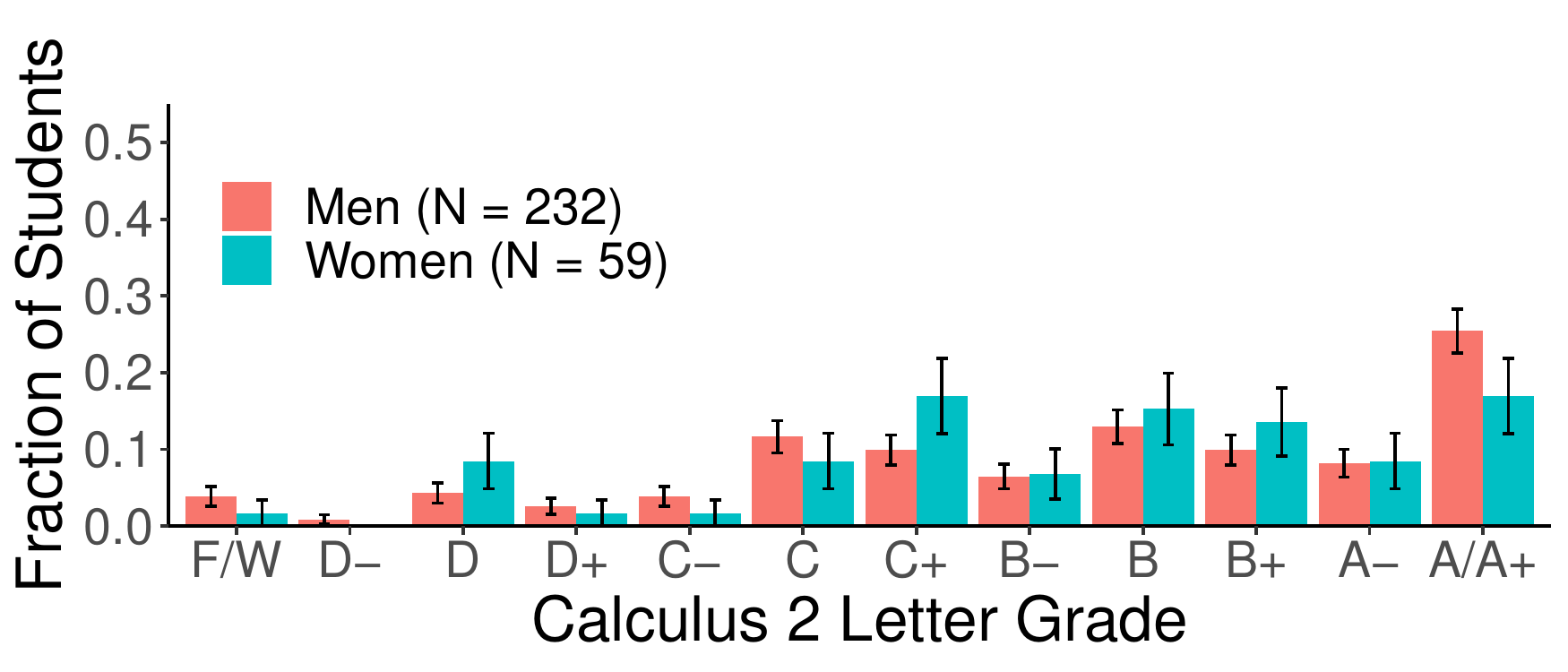}
	\end{subfigure}
		
	\begin{subfigure}{0.475\textwidth}
		\includegraphics[width=\linewidth]{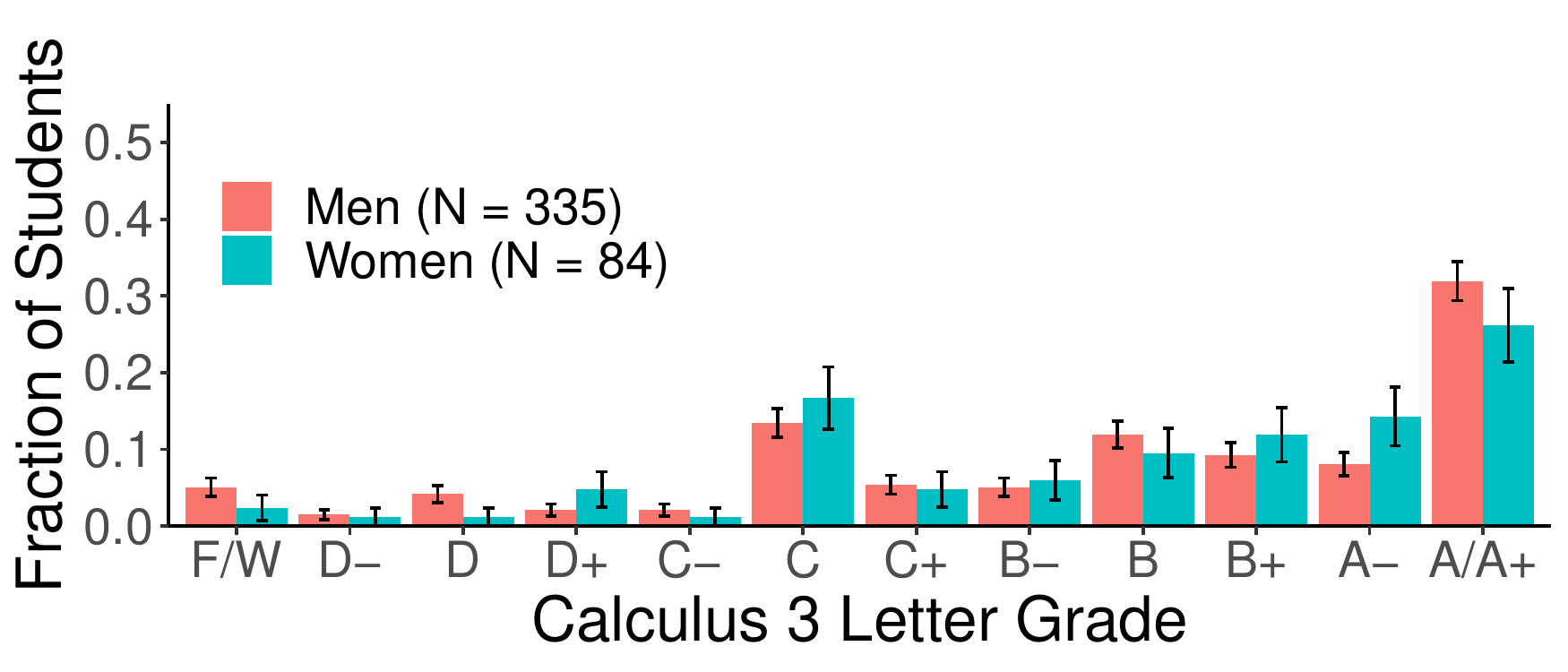}
	\end{subfigure}\hfil
	\begin{subfigure}{0.475\textwidth}
		\includegraphics[width=\linewidth]{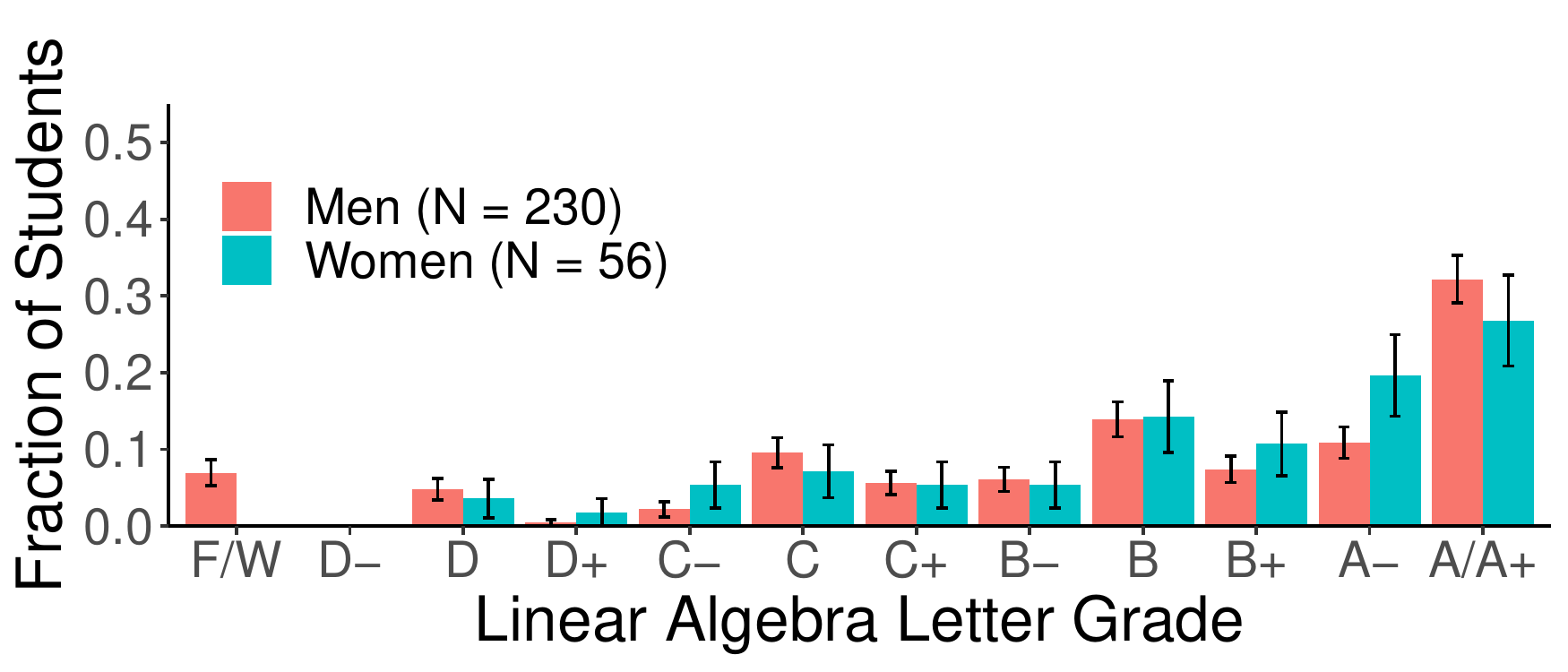}
	\end{subfigure}
	
	\begin{subfigure}{\textwidth}
		\centering
		\includegraphics[width=0.475\linewidth]{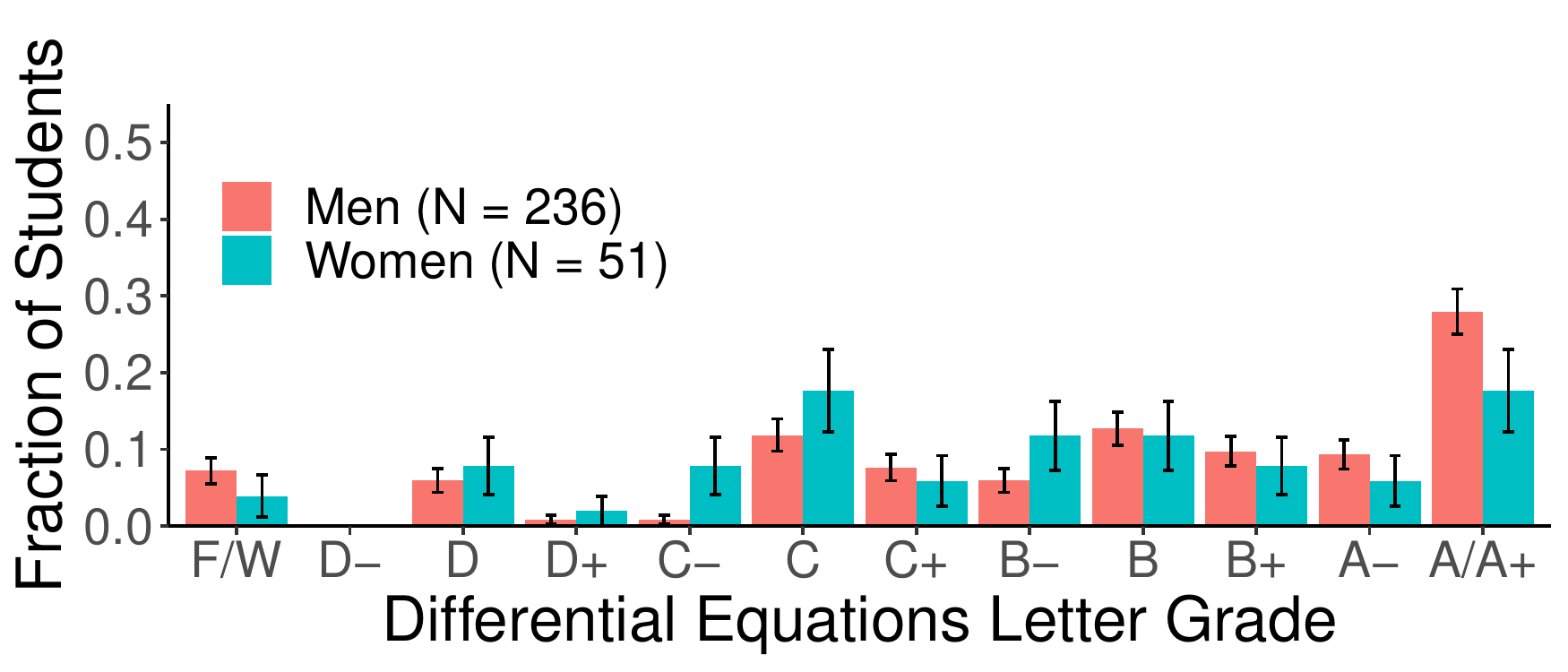}
	\end{subfigure}\hfil
	
	\label{figure_grade_dists_math}
	\caption{Grade distributions of physics majors in required mathematics lecture courses, plotted separately for men and women.
	The proportion of each gender group that earns each letter grade is plotted along with the standard error of a proportion.}
\end{figure*}

\begin{figure*}
    \centering
	\begin{subfigure}{0.475\textwidth}
		\includegraphics[width=\linewidth]{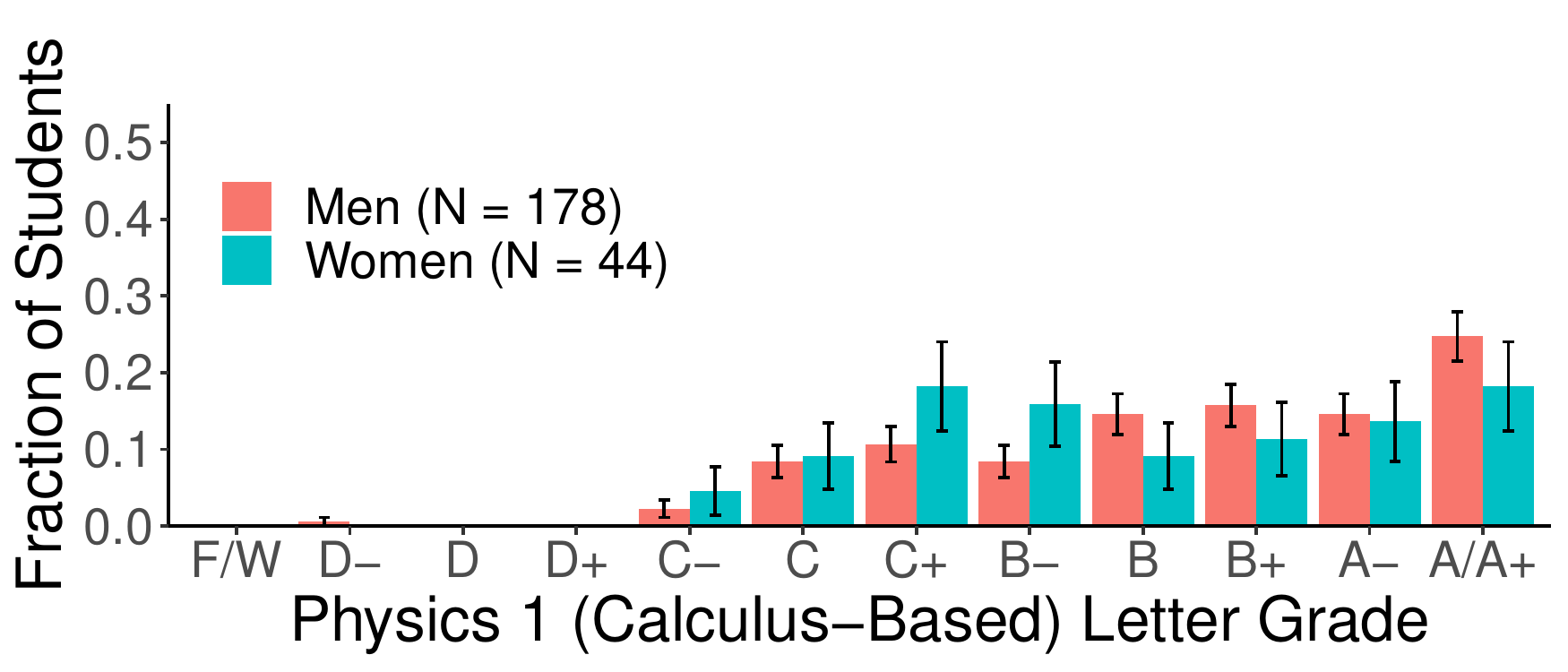}
	\end{subfigure}\hfil
	\begin{subfigure}{0.475\textwidth}
		\includegraphics[width=\linewidth]{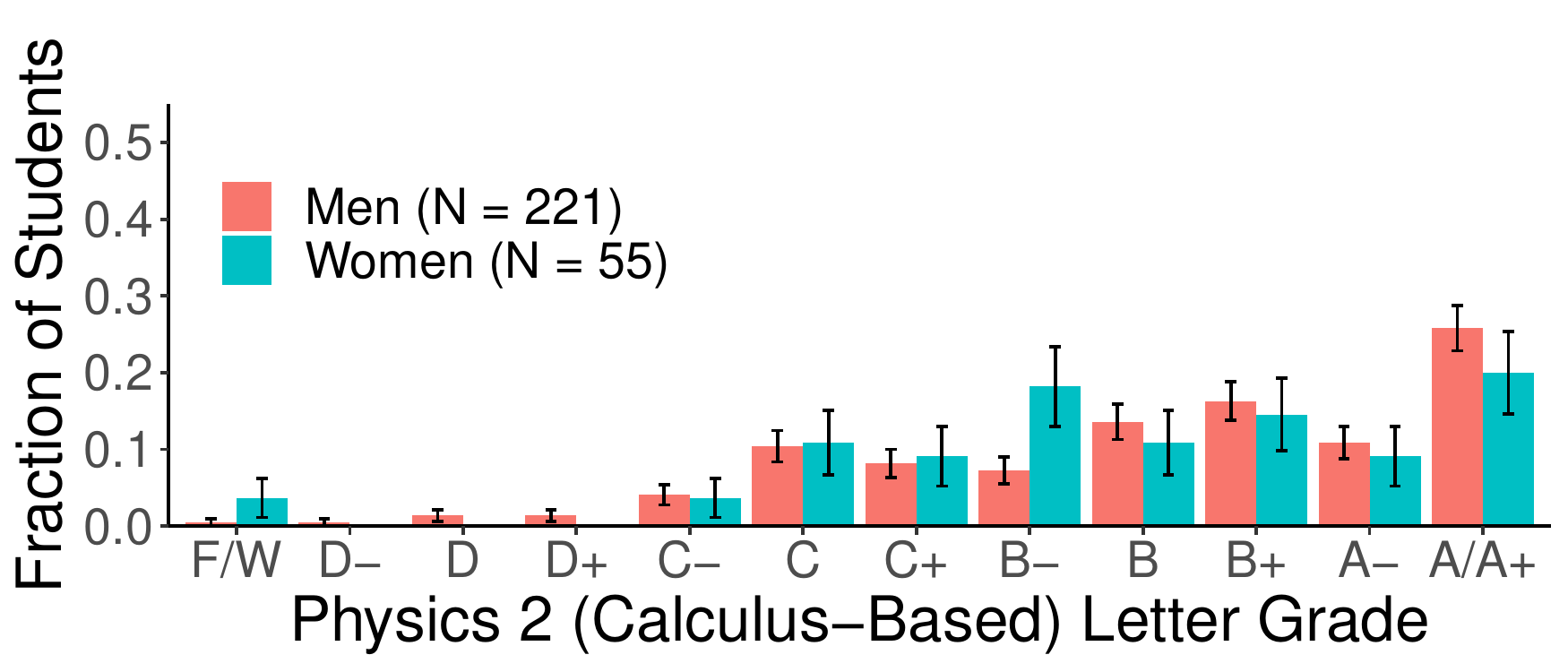}
	\end{subfigure}
		
	\begin{subfigure}{0.475\textwidth}
		\includegraphics[width=\linewidth]{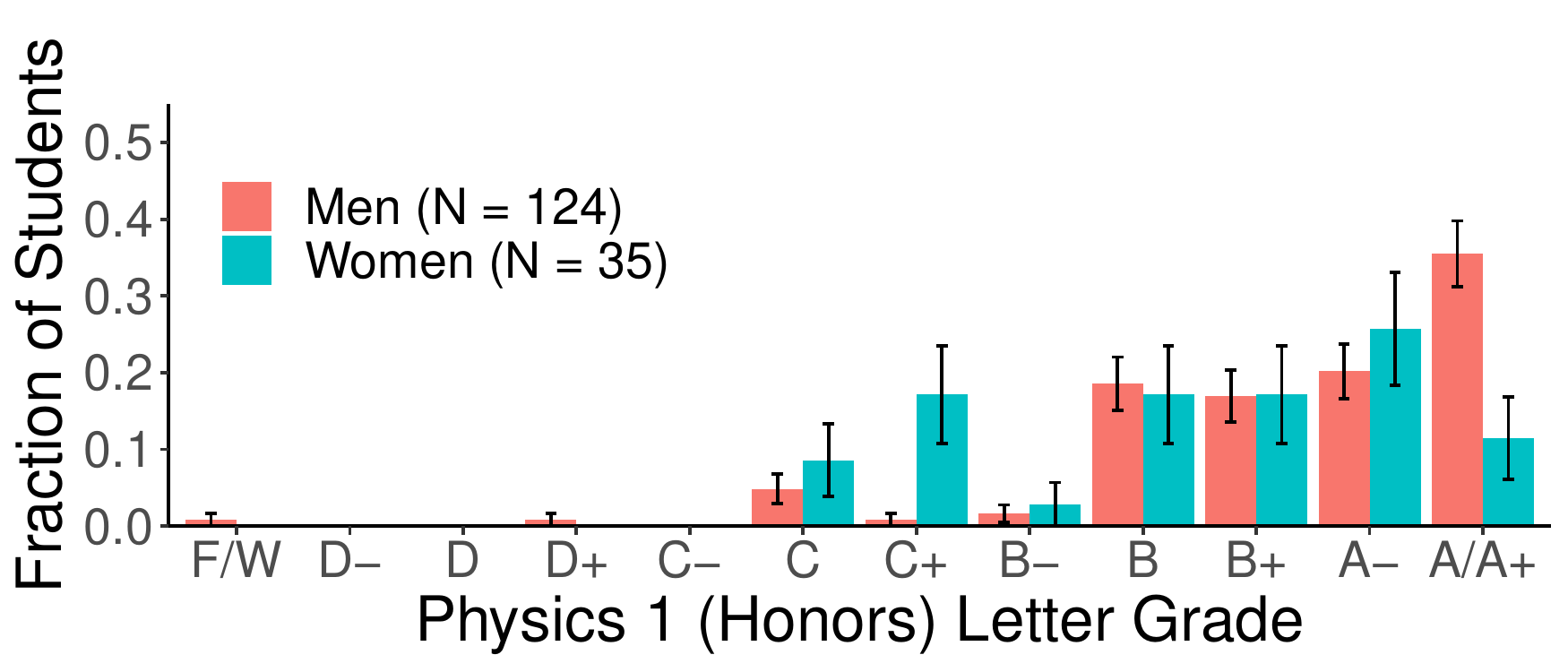}
	\end{subfigure}\hfil
	\begin{subfigure}{0.475\textwidth}
		\includegraphics[width=\linewidth]{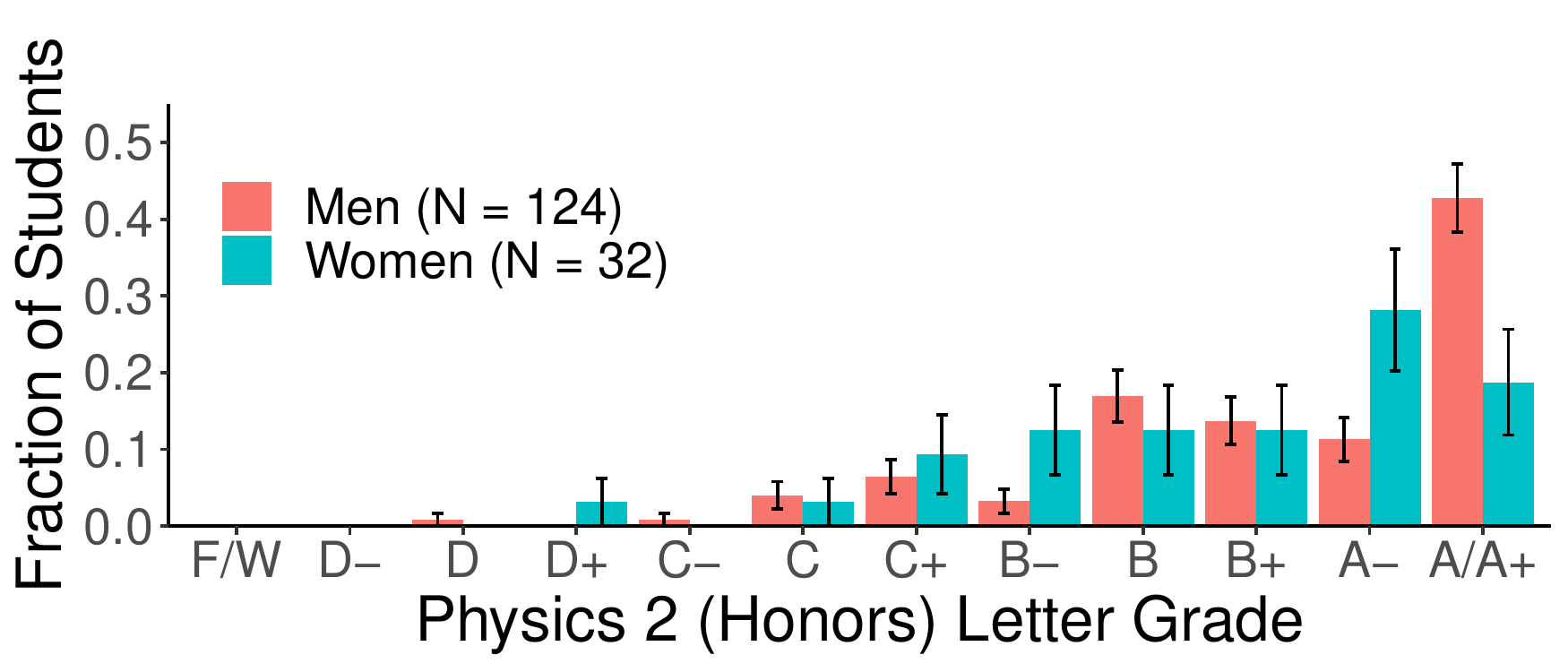}
	\end{subfigure}
	
	\begin{subfigure}{0.475\textwidth}
		\includegraphics[width=\linewidth]{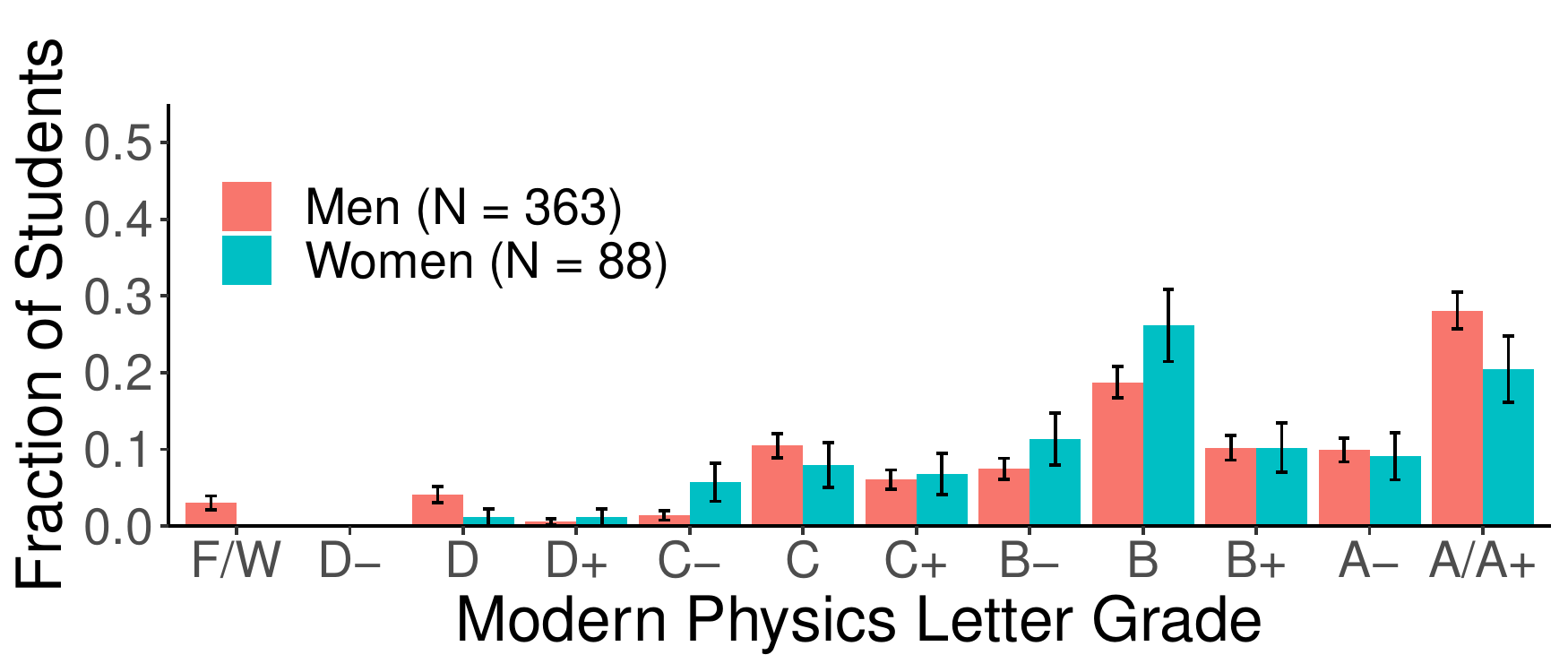}
	\end{subfigure}\hfil
	\begin{subfigure}{0.475\textwidth}
		\includegraphics[width=\linewidth]{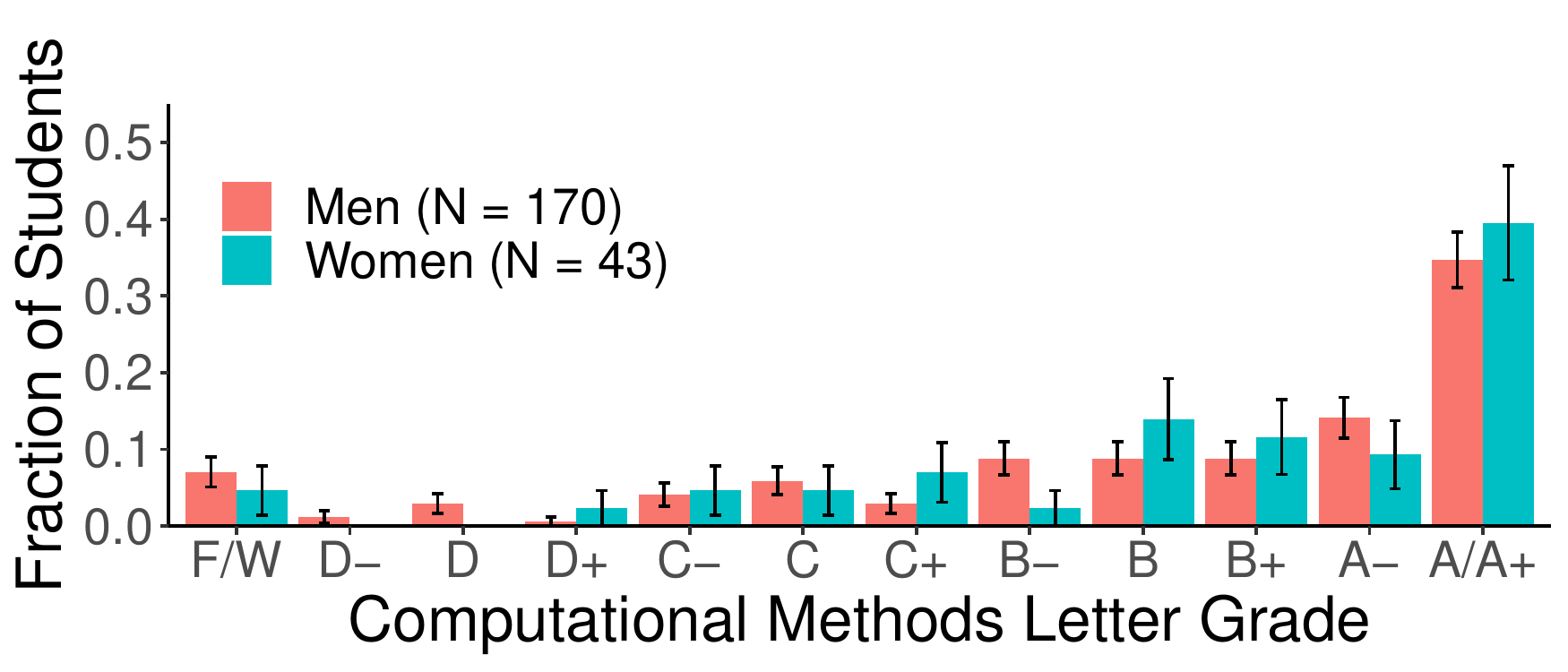}
	\end{subfigure}
		
	\begin{subfigure}{0.475\textwidth}
		\includegraphics[width=\linewidth]{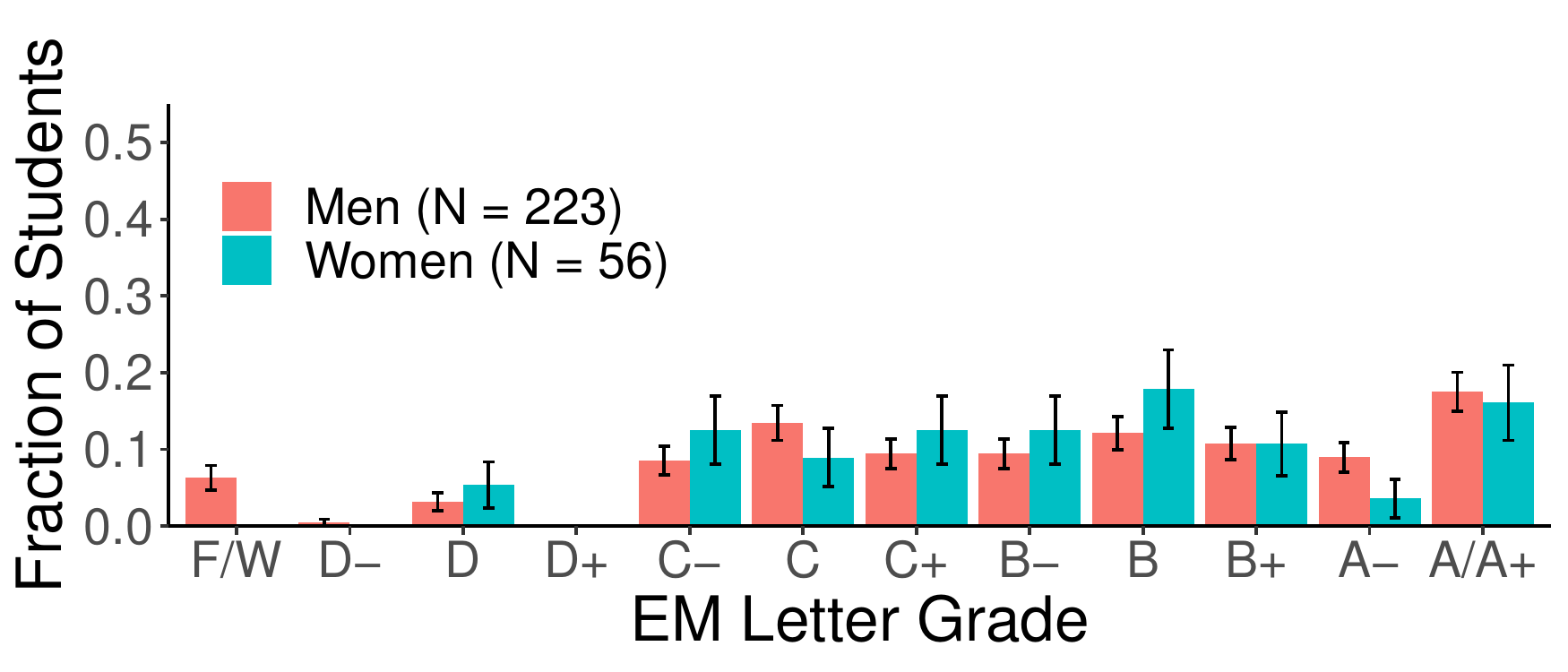}
	\end{subfigure}\hfil
	\begin{subfigure}{0.475\textwidth}
		\includegraphics[width=\linewidth]{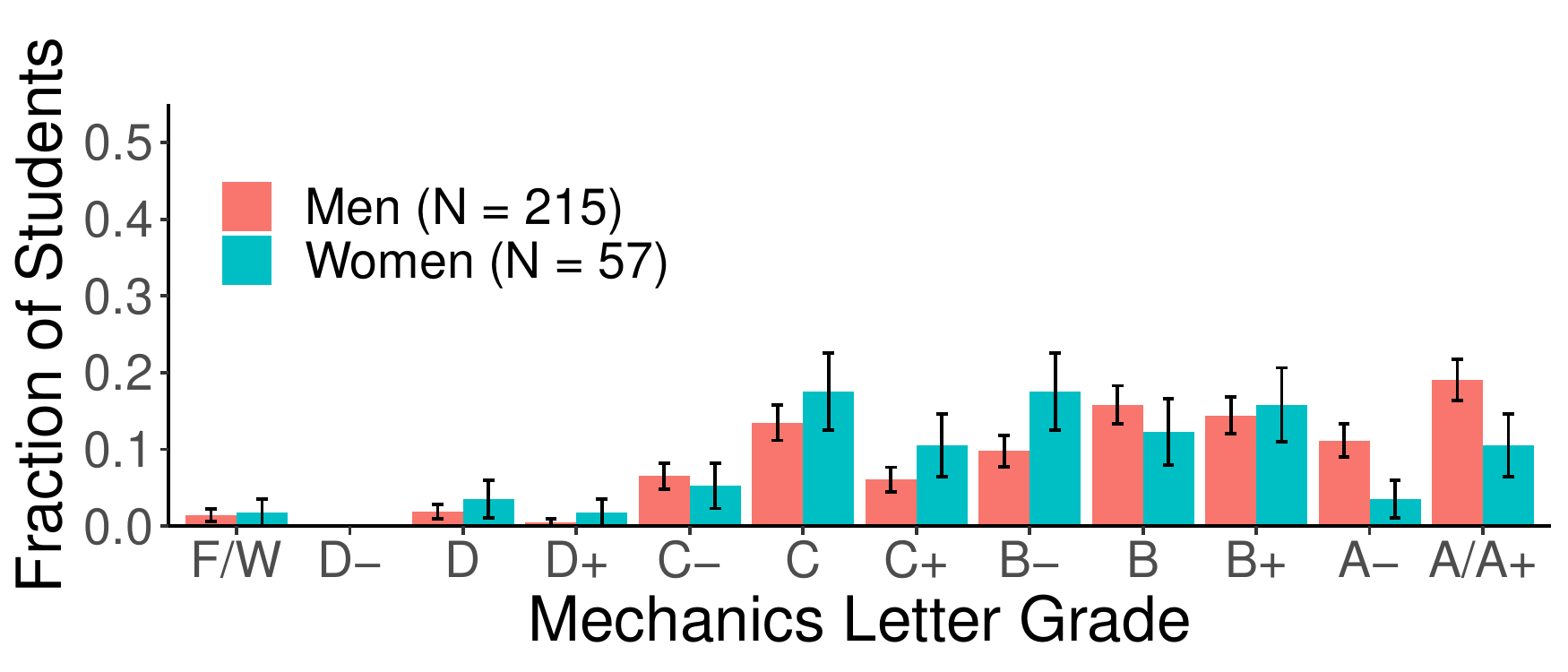}
	\end{subfigure}
	
	\begin{subfigure}{0.475\textwidth}
		\includegraphics[width=\linewidth]{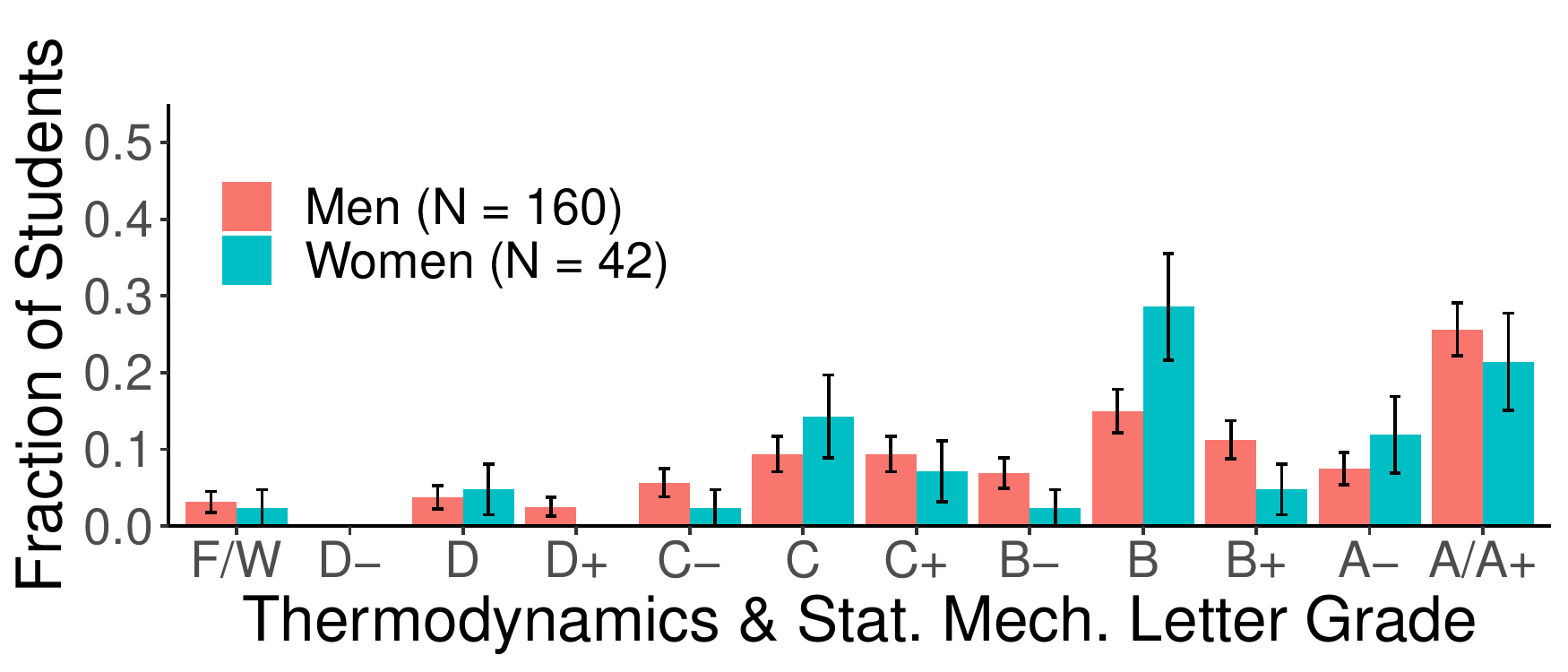}
	\end{subfigure}\hfil
	\begin{subfigure}{0.475\textwidth}
		\includegraphics[width=\linewidth]{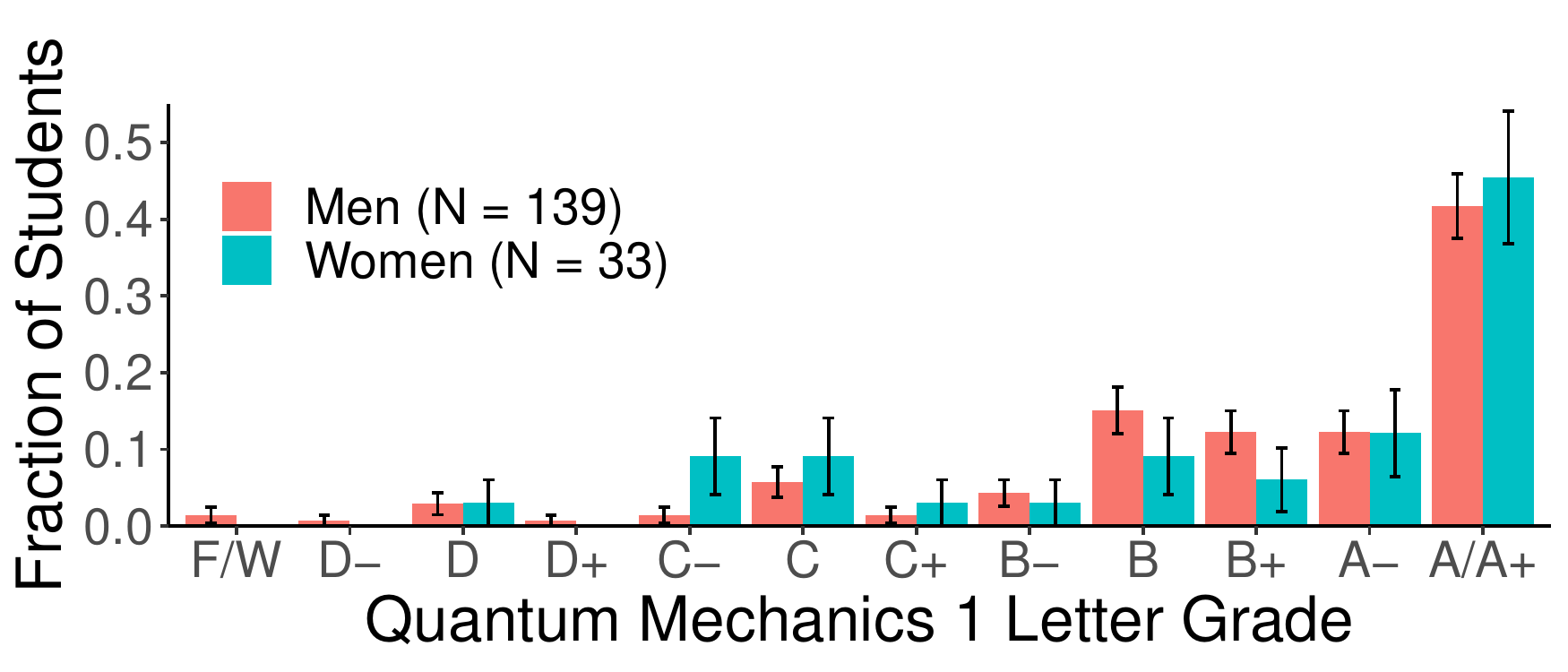}
	\end{subfigure}
	
	\label{figure_grade_dists_phys}
	\caption{Grade distributions of physics majors in required physics lecture courses, plotted separately for men and women.
	The proportion of each gender group that earns each letter grade is plotted along with the standard error of a proportion.}
\end{figure*}

\end{document}

%% file: tex/table_curriculum.tex
\begin{ruledtabular}
	\begin{tabular}{c l l}
		Typical	& Full & Shortened \\
		Term	& Course Name & Course Name \\
		\hline
		\multirow{4}{*}{1}
			& Basic Physics for Science & \multirow{2}{*}{Physics 1} \\
			& and Engineering 1			& \\
		\cline{2-3}
			& Basic Physics for Science & Honors \\
			& and Engineering 1 - Honors& Physics 1 \\
		\hline
		\multirow{4}{*}{2}
			& Basic Physics for Science & \multirow{2}{*}{Physics 2} \\
			& and Engineering 2			& \\
		\cline{2-3}
			& Basic Physics for Science & Honors \\
			& and Engineering 2 - Honors& Physics 2 \\
		\hline
		\multirow{2}{*}{3}
			& Intro to Thermodynamics, 		& Modern \\
			& Relativity and Quantum Theory & Physics \\
		\hline
		\multirow{3}{*}{5,6}
			& Computational Methods & Comp. \\
			& in Physics & Methods \\
		\cline{2-3}
			& Electricity and Magnetism & EM \\
		\hline
		\multirow{3}{*}{6,7}
			& Mechanics & Mechanics \\
		\cline{2-3}
			& Thermodynamics and & \multirow{2}{*}{Thermo} \\
			& Statistical Mechanics & \\
		\hline
		\multirow{2}{*}{7}
			& Introduction to & \multirow{2}{*}{QM 1} \\
			& Quantum Mechanics 1 & \\
		\hline
		\hline
		\multirow{2}{*}{1}
			& Analytic Geometry and & \multirow{2}{*}{Calc 1} \\
			& Calculus 1 & \\
		\hline
		\multirow{2}{*}{1, 2}
			& Analytic Geometry and & \multirow{2}{*}{Calc 2} \\
			& Calculus 2 & \\
		\hline
		\multirow{2}{*}{2, 3}
			& Analytic Geometry and & \multirow{2}{*}{Calc 3} \\
			& Calculus 3 & \\
		\hline
		\multirow{2}{*}{4}
			& Introduction to Matrices & Linear \\
			& and Linear Algebra & Algebra \\
		\hline
		\multirow{2}{*}{5}
			& Applied Differential & Diff. \\
			& Equations & Eq. \\
	\end{tabular}
\end{ruledtabular}

%% file: tex/table_introphys.tex
\begin{ruledtabular}
	\begin{tabular}{l c c c c c}
		Course	& Gender	& $N$	& $\mu$	& $\sigma$	& $d$ \\
		\hline
		\multirow{2}{*}{Physics 1}
				& F	& 64	& 2.87	& 0.92	& \multirow{2}{*}{-0.24} \\
				& M	& 238	& 3.08	& 0.83	&  \\

		\hline
		\multirow{2}{*}{Honors Physics 1}
				& F	& 71	& 3.20	& 0.63	& \multirow{2}{*}{-0.12} \\
				& M	& 221	& 3.29	& 0.83	&  \\

		\hline
		\multirow{2}{*}{Physics 2}
				& F	& 80	& 2.76	& 1.02	& \multirow{2}{*}{-0.15} \\
				& M	& 292	& 2.91	& 0.95	&  \\
		\hline
		\multirow{2}{*}{Honors Physics 2}
				& F	& 59	& 3.27	& 0.61	& \multirow{2}{*}{-0.18} \\
				& M	& 195	& 3.39	& 0.70	&  \\
	\end{tabular}
\end{ruledtabular}

%% file: tex/table_physmajors.tex
%
%

\begin{ruledtabular}
	\begin{tabular}{l c c c c c c}
		Course	& Gender	& $N$	& $\mu$	& $\sigma$	& $d$ & MANOVA \\
		\hline
		\multirow{2}{*}{Physics 1}
				& F	& 44	& 2.99	& 0.74	& \multirow{2}{*}{-0.25} &
				\multirow{8}{*}{\rotatebox[origin=c]{90}{\parbox{2.5cm}{$F(2,371)=3.13$,\\$p=0.045$}}} \\
				& M	& 178	& 3.17	& 0.73	&  & \\
		\cline{1-6}
		\multirow{2}{*}{Honors Physics 1}
				& F	& 35	& 3.13	& 0.67	& \multirow{2}{*}{-0.49} & \\
				& M	& 124	& 3.45	& 0.66	&  & \\
		\cline{1-6}
		\multirow{2}{*}{Physics 2}
				& F	& 55	& 2.93	& 0.91	& \multirow{2}{*}{-0.16} & \\
				& M	& 221	& 3.07	& 0.84	&  & \\
		\cline{1-6}
		\multirow{2}{*}{Honors Physics 2}
				& F	& 32	& 3.24	& 0.71	& \multirow{2}{*}{-0.26} & \\
				& M	& 124	& 3.42	& 0.67	&  & \\
		\hline
		\hline
		\multirow{2}{*}{Modern Physics}
				& F	& 88	& 3.03	& 0.75	& \multirow{2}{*}{0.02} &
				\multirow{12}{*}{\rotatebox[origin=c]{90}{\parbox{2.5cm}{$F(6,119)=1.52$,\\$p=0.179$}}} \\
				& M	& 363	& 3.01	& 0.98	&  & \\
		\cline{1-6}
		\multirow{2}{*}{Comp. Methods}
				& F	& 43	& 3.15	& 1.05	& \multirow{2}{*}{0.12} & \\
				& M	& 170	& 3.01	& 1.19	&  & \\
		\cline{1-6}
		\multirow{2}{*}{EM}
				& F	& 56	& 2.74	& 0.84	& \multirow{2}{*}{0.07} & \\
				& M	& 223	& 2.67	& 1.08	&  & \\
		\cline{1-6}
		\multirow{2}{*}{Mechanics}
				& F	& 57	& 2.65	& 0.84	& \multirow{2}{*}{-0.33} & \\
				& M	& 215	& 2.94	& 0.87	&  & \\
		\cline{1-6}
		\multirow{2}{*}{Thermo}
				& F	& 42	& 2.92	& 0.96	& \multirow{2}{*}{0.04} & \\
				& M	& 160	& 2.88	& 1.02	&  & \\
		\cline{1-6}
		\multirow{2}{*}{QM 1}
				& F	& 33	& 3.27	& 0.92	& \multirow{2}{*}{-0.05} & \\
				& M	& 139	& 3.31	& 0.89	&  & \\
		\hline
		\hline
		\multirow{2}{*}{Calculus 1}
				& F	& 34	& 3.15	& 0.85	& \multirow{2}{*}{0.11} &
				\multirow{10}{*}{\rotatebox[origin=c]{90}{\parbox{2.5cm}{$F(5,108)=1.07$,\\$p=0.379$}}} \\
				& M	& 149	& 3.05	& 0.99	&  & \\
		\cline{1-6}
		\multirow{2}{*}{Calculus 2}
				& F	& 59	& 2.77	& 0.97	& \multirow{2}{*}{-0.05} & \\
				& M	& 232	& 2.82	& 1.08	&  & \\
		\cline{1-6}
		\multirow{2}{*}{Calculus 3}
				& F	& 84	& 2.96	& 1.02	& \multirow{2}{*}{0.07} & \\
				& M	& 335	& 2.88	& 1.15	&  & \\
		\cline{1-6}
		\multirow{2}{*}{Linear Algebra}
				& F	& 56	& 3.15	& 0.87	& \multirow{2}{*}{0.21} & \\
				& M	& 230	& 2.93	& 1.18	&  & \\
		\cline{1-6}
		\multirow{2}{*}{Diff. Eq.}
				& F	& 51	& 2.58	& 1.06	& \multirow{2}{*}{-0.21} & \\
				& M	& 236	& 2.82	& 1.18	&  & \\
	\end{tabular}
\end{ruledtabular}